\documentclass[prb,twocolumn,showpacs]{revtex4} 

\usepackage{graphics}  
\usepackage{dcolumn}   

\newcommand{\be}{\begin{equation}} 
\newcommand{\ee}{\end{equation}}
\newcommand{\bea}{\begin{eqnarray}} 
\newcommand{\eea}{\end{eqnarray}}
\newcommand{\no}{\nonumber}

\begin{document}

\title{Energy loss mechanism for suspended micro- and nanoresonators due to the Casimir force}

\author{Andr\'e Gusso} 
\email{andre.gusso@pq.cnpq.br}
\affiliation{Departamento de Ci\^encias Exatas-EEIMVR,
Universidade Federal Fluminense, Volta Redonda, RJ, 27255-125, Brazil.}

\date{\today}

\begin{abstract}
A so far not considered energy loss mechanism in suspended micro- and nanoresonators due to noncontact acoustical energy loss is investigated theoretically. The mechanism consists on the conversion of the mechanical energy from the vibratory motion of the resonator into acoustic waves on large nearby structures, such as the substrate,  due to the coupling between the resonator and those structures resulting from the Casimir force acting over the separation gaps. Analytical expressions for the resulting quality factor $Q$ for cantilever and bridge micro- and nanoresonators in close proximity to an underlying substrate are derived and the relevance of the  mechanism is investigated, demonstrating its importance when nanometric gaps are involved. 
\end{abstract} 


\maketitle

\section{Introduction}
\label{Introduction}
High quality factor suspended micro- and nanoresonators are required in order to make practical several potential applications of such mechanical resonators  as replacements for electronic filters and reference frequency resonators as well as ultrasensitive mass, force, charge, spin, and chemical sensors 
\cite{Applications,Nguyen}. For microresonators (characterized by having at least two  dimensions in the micrometer range) the $Q$ values usually range from $\mathcal{O}(10^4)$ up to $\mathcal{O}(10^5)$. For nanoresonators (having at least two dimensions in the submicrometer range) it has usually been the case that $Q$ hardly exceeds $10^4$ (Ref. \onlinecite{SiNW}), however, more recently, nanomechanical beam resonators set to vibrate as nanostrings resulted to have $Q \sim 4 \times 10^5$ (Ref. \onlinecite{Nanostring}), and nanoresonators based on GaN nanowires vibrating in the megahertz range were reported to achieve $Q = 4 \times 10^6$ (Ref. \onlinecite{GaNNW}). As new designs and fabrication processes are created in order to overcome the known energy loss mechanisms \cite{LossMechanism}, specially clamping loss and surface defects, very high Q micro- and nanoresonators can be  expected to be available for practical applications. However, as known energy loss mechanisms are overcome increasing the quality factor, new mechanisms previously ignored can start to set new limits on $Q$.

In this work a so far not considered energy loss mechanism is investigated. This investigation is motivated by the fact that in most practical applications the resonators are expected to have their motion driven and detected electrostatically, that means capacitively, in designs involving very small gaps extending over large areas between the resonator and the electrodes. For instance, in the current practical designs of  RF MEMS filters, sub-100 nm gaps are usually required for adequate electromechanical coupling \cite{Nguyen} and while gaps as small as 20 nm where already employed \cite{Gap20nm} even smaller gaps were envisaged  as necessary for MEMS filters operation using CMOS drive voltage \cite{SmallGap}. Besides, gaps in the nanometer range are a natural consequence of the miniaturization toward NEMS filters and other devices. 

The energy loss mechanism analyzed in this work results from the coupling between the resonator and the nearby structures established across vacuum or air gaps by an attractive Casimir force.  For instance, in micro- and nanoelectromechanical resonators the nearby structures could correspond to large area electrodes built on top of the substrate and located beneath the resonator, as is usually the case for beam resonators, or the electrodes surrounding a disk resonator \cite{Nguyen}. Due to the coupling across the gap the motion of the resonator results in a time varying force on the surface of the nearby structures. This force induces the surface to oscillate at the same frequency resulting in acoustic emissions that carry away a fraction of the resonator mechanical energy. Such noncontact acoustical energy loss was considered previously in the context of tip-sample interaction in atomic force microscopy \cite{Volokitin06} due to the van der Waals force between metals and is generalized here to the interaction between the surface of micro- and nanoresonators with its surroundings mediated by the Casimir force, calculated using the full Lifshitz theory.  In the present work this mechanism is analyzed in details for the case of suspended beam resonators, considered to be located on top of a substrate. Because in practice the substrate is much larger than the micro- and nanoresonators we consider in this analysis it is modeled as a semi-space. As simplifying assumptions both the beam and the substrate are assumed to be made from a homogeneous isotropic material, and their motion is considered adiabatic (purely elastic).

\section{The energy loss mechanism}

\subsection{The Casimir force}

The Casimir force, which gives rise to the new energy loss mechanism, has  the same physical origin as the van der Waals force, resulting from the quantum fluctuations of the vacuum electromagnetic field \cite{RefCasimir}.  However, while in a simplified picture the van der Waals force can be understood as resulting from the propagation of virtual nonretarded electromagnetic waves, resulting in a short range effect, the Casimir force originates from the retarded waves that act at larger distances, extending the range of action of the quantum fluctuations. Because the Casimir force becomes relevant in the submicrometer range, its impact on the operation of MEMS and NEMS  has been receiving increasing attention \cite{GussoSNA, ReviewNEMS}. In general, this peculiar force depends on the geometry and the optical properties of the boundaries, however,  our analysis requires solely the knowledge of the negative (attractive) pressure between two semi-spaces as first derived by Lifshitz \cite{Lifshitz}. The final expression for the force is a function of the optical properties of the semi-spaces through the frequency dependent complex dielectric function.  Using the Lifshitz theory the Casimir force between semi-spaces made from materials relevant for the fabrication of micro- and nanoresonators was calculated in Ref. \onlinecite{GussoJPD}. Following this last work and the references therein, we express the Casimir force for real boundaries in terms of a correction factor to the pressure predicted for two perfectly conducting plates $P^0(d) = -\pi^2 \hbar c/(240 d^4)$, namely
\be
P(d) = -\eta(d) \frac{\pi^2}{240}\frac{\hbar c}{d^4} = -\eta(d) \frac{C}{d^4}= \eta(d) P^0(d),
\label{Pcas}
\ee  
where $d$ denotes the  gap between the surfaces, $\hbar$ the Planck constant over  $2 \pi$, $c$ is the speed of light, and the constant $C$ incorporates the constant factors in the above expression for later convenience. The factor $\eta(d)$ is usually referred to as the  finite conductivity correction factor, derived from the actual dielectric properties of the surfaces involved using the Lifshitz theory. For all known materials $\eta(d) < 1$, therefore, the pressure between two parallel surfaces made from actual materials is always smaller than the pressure between perfectly conducting plates $P^0$. The analysis presented in Ref. \onlinecite{GussoJPD} indicates the relevance of this correction factor, which is as small as 0.088 for silicon surfaces separated by a 10 nm gap, and can not be simply ignored.

\subsection{Acoustic emission and the quality factor}

Here we consider the setup were a rectangular cantilever or bridge resonator of length $l$, width $w$ and height $h$ is placed a distance $d$ above the substrate. When the resonator is set to vibrate in a given mode with time varying vertical displacement $u_n(x,t) = u_n(x) \exp(i \omega_n t)$ the gap varies according to $d-u_n(x,t)$ resulting, in the small displacement approximation, in a time varying Casimir force on the substrate. For an infinitesimal rectangular element with length $dx$ this force is
\bea
dF(x,t) &=& C \frac{\eta \biglb( d -u_n(x,t)\bigrb)}{[d-u_n(x,t)]^4} w dx \no \\
 &\approx& C \frac{\eta(d)}{d^4} w dx + 4 C \frac{\eta(d)}{d^5}w \, u_n(x) e^{i \omega_n t} dx.
\label{dF}   
\eea
The first term represents a constant force and can be ignored. It is the second time varying term proportional to $u_n(x)$ which induces a time varying displacement of substrate surface that, in its turn, results on the emission of acoustic waves with frequency $\omega_n$. The wavelength $\lambda$ of the waves produced on the substrate can be shown to be related to the dimensions of the resonator by the approximate relation  $\lambda \sim (l/h) l$, valid for the frequencies generated by the first three modes. Because for most practical devices $l/h \gtrsim 10$ and $w < l$,  $\lambda$  is large compared with the lateral dimensions of the source, therefore, justifying the use of the point source approximation. In this approximation the details on the force distribution over the source are not of fundamental importance for the calculation of the irradiated acoustic power. However, some aspects of the force distribution must be taken into account as we do next.

\begin{figure}
\resizebox{6.5 cm}{8.86 cm}{
\includegraphics{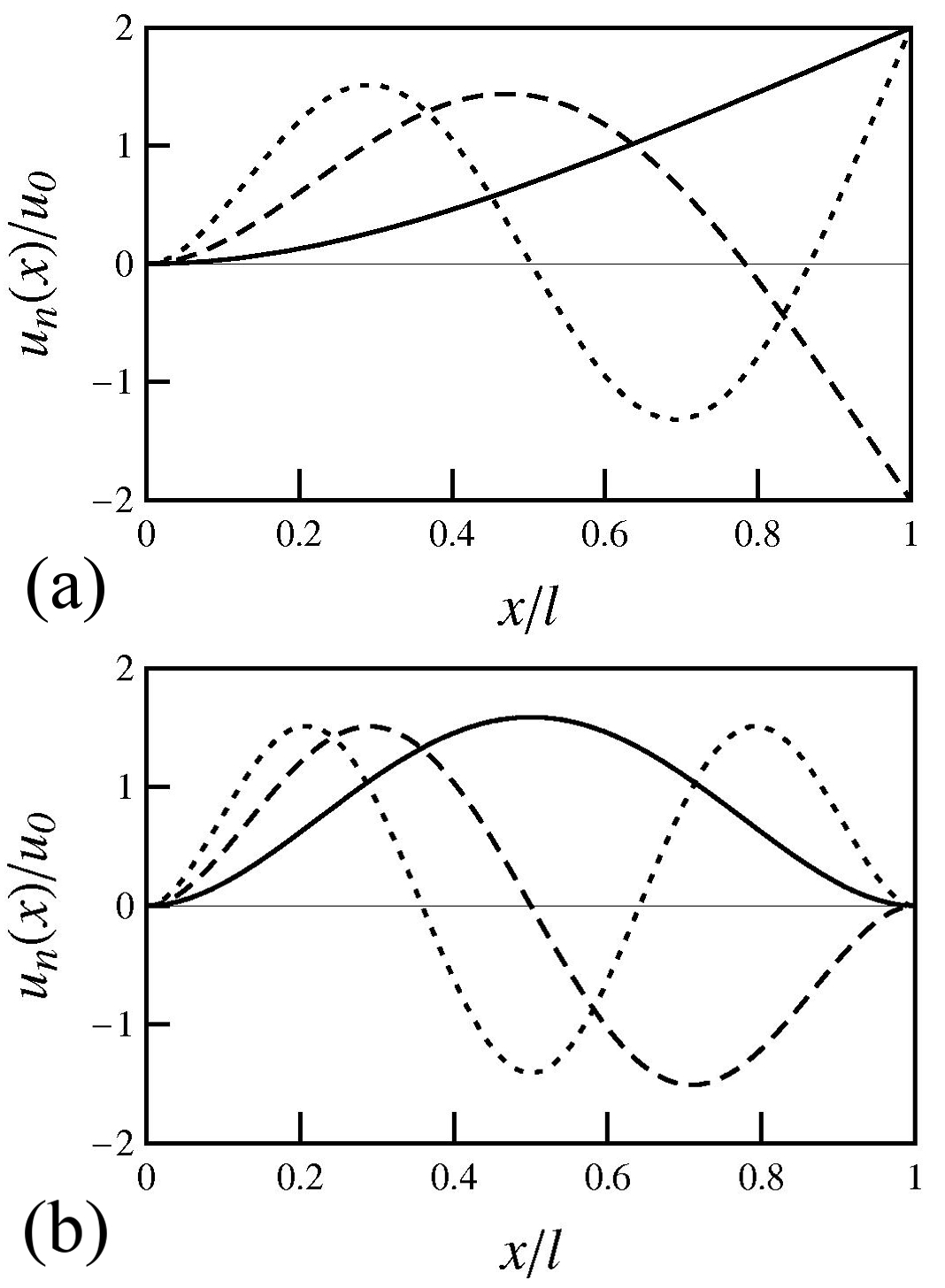}}
\caption{Mode shapes for the three lowest frequencies $n=1$ (continuous), 2 (dashed), and 3 (dotted) for (a) cantilever and (b) bridge resonators.    \label{modes}}
\end{figure}

Firstly, we note that  in most practical applications of beam micro- and nanoresonators the ratio $h/l$ is sufficiently small to the Euler-Bernoulli beam theory to apply \cite{Graff}, at least approximately. In this case the resulting mode shapes are given by a general expression of the form
\bea
u_n(x) &=& u_0 \{ \cosh(\kappa_n x/l) -\cos(\kappa_n x/l) \no \\
&+& \chi_n \left[\sinh(\kappa_n x/l) -\sin(\kappa_n x/l) \right] \}
\eea
where for the first three modes of the cantilever(bridge) we have $\kappa_1 = 1.8751(4.7300), \kappa_2 = 4.6941(7.8532), \kappa_3 = 7.8548(10.996)$, and $\chi_1 = -0.7341(-0.9825), \chi_2 = -1.0185(-1.0008)$, and $\chi_3 = -0.9992(-0.9999)$. In Fig. \ref{modes},  $u_n(x)$ for the first three modes of cantilever and bridge are presented. As seen from the second term in Eq. (\ref{dF}) the amplitude of the time varying force on the surface varies in position following  $u_n(x)$. Therefore, all the points of the source region on the surface are in phase for the first mode of both the cantilever and bridge, and both sources can be held as acoustic monopoles. For the second mode of the bridge,  points to the left and to the right of the mid point vibrate out of phase by $180^o$ and the source can be treated as an acoustic dipole. Before discussing how the acoustic emissions produced by the other vibrating modes can be held, let us consider the acoustic emission by the monopoles and dipoles on the surface. 

Miller and Pursey \cite{MP}  were the first to derive from the elasticity theory an expression for the acoustic energy emitted by a point source vibrating normal to the surface of a semi-space.  For an harmonically varying force of the form $F(t) = F \exp(i \omega_n t)$ they employed the admittance method obtaining \cite{MP}
\be
\Pi^m = \frac{1}{4 \pi} \frac{\sqrt{\rho^s \, C^s_{11}}}{{C^s_{44}}^2} \omega_n^2 \,F^2 \,\phi_m(\gamma)
\label{powermonopole}
\ee
where $\phi_m(\gamma)$ corresponds to 
\be
\phi_m(\gamma) = Im \left[ \int_0^\infty \frac{p \sqrt{p^2-1}}{F_0(p, \gamma)} \, dp \right]
\label{IntegralMP}
\ee
with
\be
F_0(p, \gamma) = (2 p^2 -\gamma^2)^2-4 p^2 \sqrt{p^2-1}\sqrt{p^2-\gamma^2 },
\label{denominator} 
\ee 
and $\gamma = \sqrt{C^s_{11}/C^s_{44}}$, $\rho^s$ is the density, $C^s_{11}$ and $C^s_{44}$ are the elastic stiffness coefficients that characterize the isotropic material of the substrate, as indicated by  the superscript $s$. Several authors have rederived Eq. (\ref{powermonopole}) since the now classical work of Miller and Pursey, however, differying in the definition\cite{Judge} or on the evaluation\cite{Volokitin06,Hao09} of $\phi_m(\gamma)$.  In the original work\cite{MP} the integral was evaluated numerically taking into account the branch-points $p=1, \gamma$, the principal value of the radicals, and the only physically relevant pole satisfying the condition $p>\gamma$. Following the prescriptions given by Miller and Pursey in the Section 7 of Ref. \onlinecite{MP}(a) we derive another representation suitable for numerical evaluation
\bea
\phi_m(\gamma) &=& \int_0^1 \frac{p \sqrt{1-p^2}}{(2p^2 -\gamma^2)^2+4p^2\sqrt{1-p^2}\sqrt{\gamma^2-p^2}}dp \no \\ 
&+& \int_1^\gamma \frac{4 p^3(p^2-1) \sqrt{\gamma^2-p^2}}{(2 p^2-\gamma^2)^4+16p^4(p^2-1)(\gamma^2-p^2)}dp \no \\ 
&-& \pi \frac{p_r \sqrt{p_r^2-1}}{F_0^\prime(p_r,\gamma)},
\label{Im}
\eea  
where the last term corresponds to the contribution from a clockwise indentation around the pole at $p=p_r$, determined as the root of $F_0(p,\gamma)=0$ satisfying $p_r > \gamma$. The above representation was checked to reproduce the numerical result reported by Miller and Pursey for $\gamma = \sqrt{3}$, $\phi_m(\sqrt{3}) = 0.537$, and to differ only sligthly from the result reported by Hunter\cite{Hunter57} for $\gamma =2$, $\phi_m(2) = 0.415$, in which case we obtain $\phi_m(2)=0.409$, a difference that may be due to numerical precision. It is worth to note that in Ref. \onlinecite{Judge}, in spite of the fact that the authors base their analysis on the work of Miller and Pursey, the expression for the acoustic power $\Pi^m$  contains an integral that, while similar to that in Eq. (\ref{IntegralMP}), leads to significant discrepancies in the numerical results, underestimating the emitted acoustic power by as much as a factor of 10. In a recent analysis of the support (or clamping) loss in micromechanical resonators\cite{Hao09}, which also  follows the work of Miller and Pursey, exactly the same expression for $\Pi^m$ given in Eq. (\ref{powermonopole}) was reported. However, the numerical evaluation of $\phi_m(\gamma)$ did not take into account the contribution from the pole at $p_r$, also leading to a significant underestimate of acoustic emissions. The neglecting of the contribution from the pole corresponds, physically, to neglecting the contribution of the Rayleigh surface waves which are responsible for carrying away the major fraction of the acoustic energy\cite{MP}. In Ref. \onlinecite{Volokitin06} the acoustic emission by an harmonically varying normal force applied to the surface of a semi-space was reconsidered. This work provides a coefficient of friction $\Gamma_\bot = \xi_\bot F^2/(4 \pi \rho^s c_t^3)$, where $c_t=\sqrt{C^s_{44}/\rho^s}$ is the transverse sound velocity, and $\xi_\bot$ corresponds to the sum of three terms similar to those in Eq. (\ref{Im}). From $\Gamma_\bot$ we can derive the emitted acoustic power defined as\cite{Volokitin06}, $\Pi = \Gamma_\bot 2 \omega^2 u_0^2$ were, in the notation of Ref. \onlinecite{Volokitin06}, $u_0$ denotes half the amplitude of the vertical motion which was given as the sum of a complex amplitude proportional to $u_0$ plus its conjugate. The resulting expression for $\Pi^m$ is identical to that  in Eq. (\ref{powermonopole}) with $\phi_m(\gamma)$ replaced by $\xi_\bot/(2 \gamma)$. By means of an adequate change of variables, $\phi_m(\gamma)$ can be made equal to  $\xi_\bot/(2 \gamma)$ , except for the limits of integration  of the two integrals in Eq. (\ref{Im}). Therefore, the emitted acoustic power calculated using the expressions provided in Ref. \onlinecite{Volokitin06} does not match $\Pi^m$ determined  by Miller and Pursey. However, the results can be made to coincide if the limits defined in the two integrals in $\xi_\bot$ are taken squared, in which case they become the same as the limits in Eq. (\ref{Im}) after the proper change of variables, indicating that $\xi_\bot$ should be corrected in this manner.

Specializing to the case of a suspended resonator vibrating transversally in the mode $u_n(x)$, the time varying contribution from the total applied force on the surface is
\bea
F(t) &=& F \exp(i \omega_n t) \no \\
&=& 4 C \eta(d) d^{-5} w \int_0^l u_n(x) dx \exp(i \omega_n t) ,
\label{verticalforce}
\eea
Therefore, the energy lost per cycle is $\Delta U_n = \Pi/f_n = 2 \pi \Pi/ \omega_n$. The resulting quality factor $Q$ is a measure of the ratio of the vibrational energy of  the resonator $U_n$ to the energy lost, namely, $Q= 2 \pi U_n/\Delta U_n = \omega_n U_n/\Pi$. The vibrational energy  for both  cantilevers and bridges is $U_n = h w l \rho^r \omega_n^2 u_0/2$ while the mode frequency is $\omega_n^2 = \kappa_n^4 E^r h^2/(12 \rho^r l^4)$, implying  that for the modes considered as acoustic monopoles
\be
Q^m = \frac{\pi}{16 \sqrt{3}} \frac{\kappa_n^2}{I_{u_n}^2} \frac{1}{C^2 \phi_m(\gamma)} {C_{44}^s}^2 \left( \frac{E^r \rho^r}{C_{11}^s \rho^s} \right)^{1/2} \frac{h^2}{w l^3} \frac{d^{10}}{\eta(d)^2},
\label{monopole}
\ee
where $I_{u_n} = \int_0^l u_n(x) dx/(u_0 l)$, $E^r$ denotes the Young modulus, and $\rho^r$ the density of the resonator as indicated by the superscript $r$. The result expressed in Eq. (\ref{monopole}) reveals that $Q^m$ is a fast varying  function of the gap distance showing an explicit dependence that goes as $d^{10}$. However, in order to determine the actual dependence of $Q^m$ on $d$ we have to take into account the term $\eta(d)$. From the results presented in Ref. \onlinecite{GussoJPD} it is generally the case for conductors and semiconductors that $\eta(d) \propto d^\alpha$ with $\alpha$ increasing almost linearly from approximately 0.65 for $d$ equal to 15 nm to values close to one at 1 nm. Therefore, in this particular range of distances, $Q^m$ has an exponent for $d$ varying from 8 up to a maximum of approximately 8.7.  The dependence on the geometrical parameters $h, w$, and $l$ indicates that the new energy loss mechanism is more relevant for thin, wide, and long structures. It can also be inferred that $Q^m$ is smaller for soft materials in the substrate due to the dependence on ${C_{44}^s}^2$.

We turn now to the analysis of a dipolar excitation at the surface of the substrate. In this case there is no net vertical force, instead there is a net bending moment $M$ which, due to the symmetry of the second mode for the bridge, results to be
\be
M = 8 C \frac{\eta(d)}{d^5} w \int_0^{l/2} u_2(x) \left( \frac{l}{2} - x \right) dx.
\ee  
This bending moment causes the surface to twist. For small sources, we can use the average twisting angle  calculated by Bycroft\cite{Bycroft} for an harmonically varying bending moment $M(t)= M \exp(i \omega_n t)$ distributed over a circular region, Eq. (191) of Ref. \onlinecite{Bycroft}. In the limit of small radius over wavelength ratio the amplitude of the angle is
\be
\theta =  \frac{M \, k^3 }{4 \pi \gamma C^s_{44}} \phi^\prime_d(\gamma) ,
\ee
where
\be
\phi^\prime_d(\gamma)=\int_0^\infty \frac{p^3 \sqrt{p^2-1}}{F_0(p,\gamma)} \, dp ,
\ee
and $k=2 \pi/\lambda^s=\omega_n \sqrt{\rho^s/C^s_{44}}$. We implicitly incorporate into the integral the explicit contribution of the physically allowed pole at $p_r$ introduced by Bycroft. This last author also introduces the  same explicit contribution into the expression for the average vertical displacement due to an harmonically varying normal force when compared to the result obtained by Miller and Pursey\cite{MP}. With this definition for the integral we can proceed to obtain a representation  suitable for numerical evaluation following the same procedure  adopted for $\phi_m(\gamma)$.  As we argue next, in order to determine the emitted acoustic power only the evaluation of the imaginary part of  $\phi^\prime_d(\gamma)$ is required, which can be written as
\bea
\phi_d(\gamma) &=& Im[\phi^\prime_d(\gamma)] \no \\
&=& \int_0^1 \frac{p^3 \sqrt{1-p^2}}{(2p^2 -\gamma^2)^2+4p^2\sqrt{1-p^2}\sqrt{\gamma^2-p^2}}dp \no \\ 
&+& \int_1^\gamma \frac{4 p^5(p^2-1) \sqrt{\gamma^2-p^2}}{(2 p^2-\gamma^2)^4+16p^4(p^2-1)(\gamma^2-p^2)}dp \no \\ 
&-& \pi \frac{p_r^3 \sqrt{p_r^2-1}}{F_0^\prime(p_r,\gamma)}.
\label{Id}
\eea
In the complex notation the real and imaginary components of the displacement correspond to the in-phase and out-of-phase components relative to the applied force or moment. Only the out-of-phase component of $\theta$ contributes to the time averaged  acoustic power through $\Pi = \langle Re[M(t)] Re[\dot{\theta}(t)]\rangle = Re[M \times \dot{\theta}]/2$, where $\langle \; \rangle$ denotes the time average, $M$ the real amplitude from $M(t)$, and $\dot{\theta}=i \omega_n \theta$ the complex amplitude of the angular velocity. We note that we could have used the analogous definition for the average emitted acoustic power for the normal point source obtaining the same result as in Eq. (\ref{powermonopole}). In this case $\Pi = \langle Re[F(t)] Re[\dot{z}(t)] \rangle$, where $z(t) = z \exp(i \omega_n t)$, and $z$ denotes the complex average vertical displacement given by Eq. (129) of Ref. \onlinecite{MP}(a). In the case of the dipolar source the resulting emitted power is 
\be
\Pi^d =  \frac{1}{{8 \pi C_{44}^s}^2}\left( \frac{{\rho^s}^3}{C^s_{11}}\right)^{1/2}\omega_n^4 M^2 \phi_d(\gamma)
\label{Powerdipole}
\ee
As for $\Pi^m$ it is worth to compare this result for $\Pi^d$ with those found in the literature. Compared to the results presented in Ref. \onlinecite{Judge}, also based upon the work of Bycroft, the same difference is found concerning the definition of the denominator in the integral $\phi^\prime(d)$. Our result can also be compared with a derivation for the acoustic energy loss due to an AFM tip vibrating parallel to the surface of a plane substrate\cite{Volokitin06}. This is an analogous situation because the harmonic horizontal displacement of the force can be interpreted as resulting into a time-varying harmonic torque about an horizontal axis perpendicular to the direction of the tip vibration. Considering the definition of the 
horizontal displacement given in Ref. \onlinecite{Volokitin06}, which has amplitude $2 u_0$, a point force displacing horizontally results in a torque $M(t) = 2 u_0 F \cos(\omega_n t)= M \cos(\omega_n t)$. From the given expression for the friction coefficient (see the Appendix B of Ref. \onlinecite{Volokitin06}) 
\be 
\Gamma_\| = \frac{\xi_\|}{8 \pi} \frac{\omega_n^2}{\rho c_t^5} F^2,
\ee
the emitted power $\Pi = \Gamma_\| 2 \omega^2 u_0^2$ results to be the same as that in Eq. (\ref{Powerdipole}) with $\phi_d(\gamma)$ replaced by $(\gamma/2) \xi_\|$, where $\xi_\|$ is an expression similar to Eq. (\ref{Id}). As for the monopole case, $\phi_d(\gamma)$ can be made to coincide with $(\gamma/2) \xi_\|$ after an adequate change of variables except for the limits of integration. The two results for $\Pi^d$ can be made identical if the limits of integration in $\xi_\|$ are taken squared.

From Eq. (\ref{Powerdipole}) we can follow the same procedure as for the monopole in order to calculate the  quality factor for the second mode of the bridge which results to be
\be
Q^d = \frac{1.222}{C^2 \phi_d(\gamma)} \, {C_{44}^s}^2 \left( \frac{C_{11}^s {\rho^r}^3}{E^r {\rho^s}^3} \right)^{1/2} \frac{1}{w l} \frac{d^{10}}{\eta(d)^2} .
\label{dipole}
\ee
Compared to the $Q^m$ calculated for the first mode of the bridge, $Q^d$ is larger by roughly  a factor $(l/h)^2$. This factor is exactly what would be expected from the dipolar nature of the source. The power irradiated by an acoustic dipole where the two sources are separated by a distance $D=l/2$, as is approximately the case here, is proportional to $(D/\lambda^s)^2=(l/ 2 \lambda^s)^2$ times the energy irradiated by a single monopole with the same strength \cite{Fahy}. Therefore, as $\lambda^s \propto l^2/h$ it results that $Q^d \propto (l/h)^2 Q^m$, as noted above. In fact, this relation between $Q^d$ and $Q^m$ is generally valid and because in most of the bridge and cantilever resonators found in the literature the ratio $l/h$ is close to or larger than 10, the energy loss tends to be larger for resonators vibrating in such a way as to produce a net vertical force on the surface of the substrate as compared to a net bending moment.

\subsection{Beyond acoustic monopoles and dipoles}

The expressions for $Q^m$ and $Q^d$ as derived above are strictly valid for vibrational modes resulting in acoustic monopoles and dipoles, respectively. However, we can expect that Es. (\ref{monopole}) and (\ref{dipole})  provide approximate results for slightly more complex vibrations of the resonators whenever net vertical forces or net bending moments are the prevailing disturbances acting on the surface of the substrate. This fact allow us to extend the results for some higher order vibrational modes. The estimate of $Q$ for higher order modes is important because the use of such modes in practical devices is becoming an alternative as a means to achieve high frequency operation, specially in the UHF range \cite{Nguyen, SmallGap}. Focusing on the first three modes of the cantilever and bridge resonators, we can firstly argue that the second mode of the cantilever is going to lose energy predominantly as a monopole due to the net vertical force produced on the substrate, while the portion of the cantilever vibrating out of phase emits energy as a dipole at a much smaller rate. 

In order to clarify this argument, we note that the net vertical force given by Eq. (\ref{verticalforce}) is the same for every mode, the difference coming from the integral over $u_n(x)$. As noted after Eq. (\ref{monopole}) this integral can be written as $I_{u_n} u_0 l$ were $I_{u_n} = 0.783$ for the first mode of the cantilever and equal to 0.434 for the second mode. Therefore, the net vertical force produced by the second mode is large, comparable to the force for the first mode, resulting in acoustic emissions that exceed any dipolar emissions produced by the small portion of the source close to the free end of the cantilever. In its turn,  the third mode of the cantilever produces a surface force distribution that is close to the dipolar source produced by the second mode of the bridge. However, in this case, due to the lack of symmetry of this mode there results both a net vertical force and net bending moment. The vertical force is reduced compared to the lower order modes being, in this case, proportional to $I_{u_n}=0.254$. The bending moment at the more characteristically dipolar portion of the source, limited to the left of the second node at $x=0.868 \, l$, is close to that found for the second mode of the bridge. As a consequence, for two limiting cases where the ratio $l/h$ is sufficiently small (large) that  the predicted energy loss due to the bending of the surface is much larger (smaller) than that  due to the net vertical displacement the quality factor can be estimated using Eq. (\ref{dipole}) [Eq. (\ref{monopole})]. Finally,  the third mode of the bridge can be treated approximately as an acoustic tripole, a source comprised of three monopoles, two inphase and one out of phase by $180^o$. In this case the power emitted is that produced by a single monopole that causes a net vertical force  proportional to $I_{u_n}= 0.364$.  

\begin{figure}
\resizebox{8 cm}{3.92 cm}{
\includegraphics{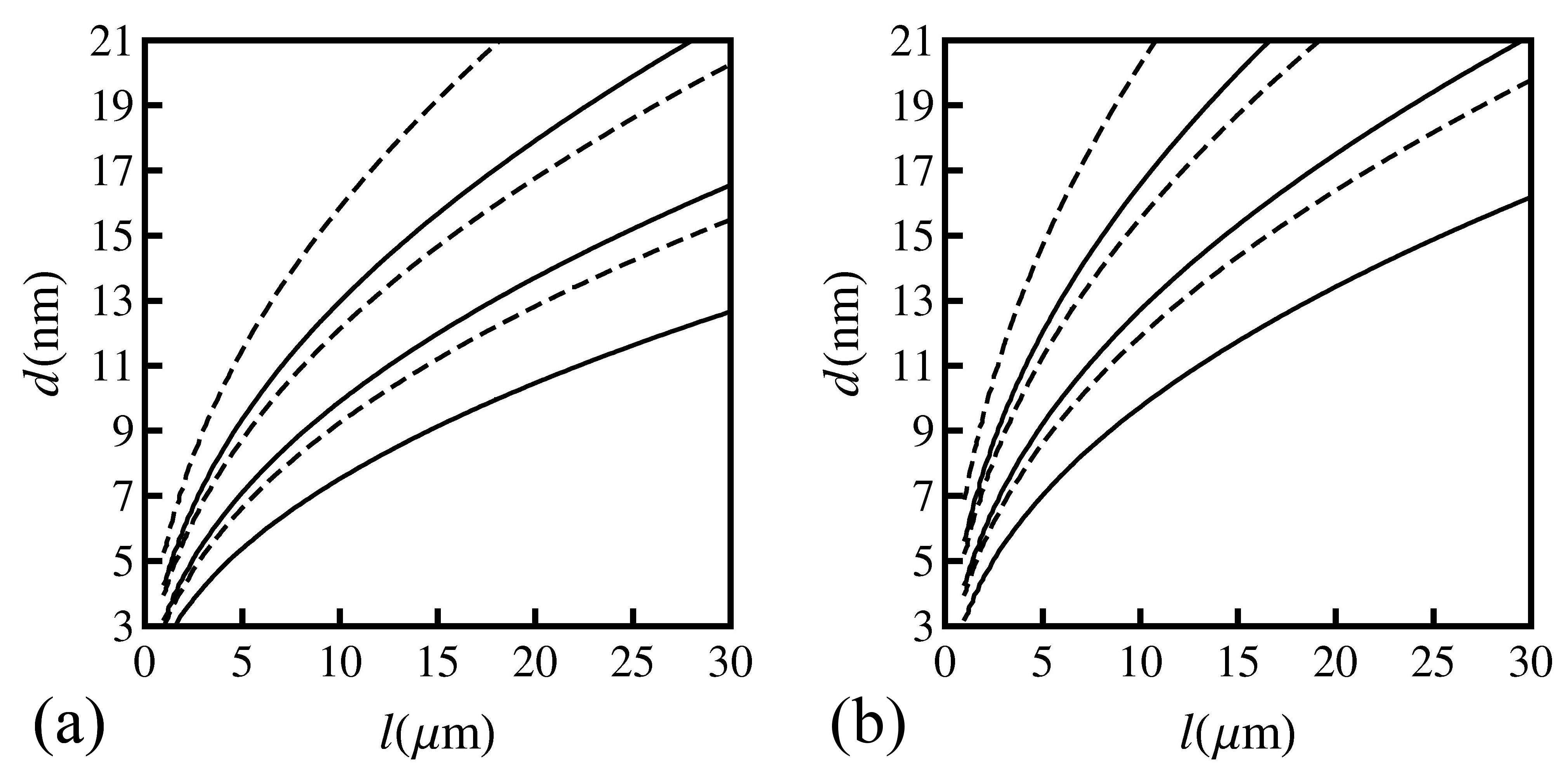}}
\caption{Contourplot for the quality factor $Q$ as a function of the gap $d$ and length $l$ of the resonator for a fixed ratio $l/w = 5$, and thickness $h= 0.1 \, \mu$m. Resonator and substrate made from (a) polysilicon and (b) gold; continuous(dashed) lines are for bridge(cantilever) resonators;  the contours are for $Q = 10^4$ (bottom curve), $Q=10^5$ (middle curve), and $Q=10^6$ (upper curve).      \label{contour}}
\end{figure}

From the analysis presented so far, some general trends  on $Q$ for higher order modes can be advanced. In general, the acoustic sources at the substrate surface are going to be $n$-poles, corresponding to the $n$ antinodes, each pole  having approximately the same  shape, and consequently intensity, along the resonator. For $n$ odd there is a net vertical force that decreases significantly for $n \geq 3$ resulting in a very large ratio $k_n/I_{u_n}$ and, therefore, on the increase of $Q$ with $n$ [see Eq. (\ref{monopole})]. For $n$ even, there is no net vertical force and an increase of $Q$ with $n$, proportional to $(D/\lambda^s)^{-n} \sim (l/h)^n$, is expected based on general results for the acoustic emission by multipoles \cite{Fahy}. 

\section{Results and conclusions}

In order to illustrate the relevance of the new energy loss mechanism we present in  Fig. \ref{contour} contourplots for different values of $Q$. We consider cantilever and bridge resonators made from polysilicon and gold. Polysilicon is chosen because most MEMS and NEMS are made based mostly on this material, and results based on it are representative of other forms of silicon and  semiconductors like gallium arsenide and germanium, due to their similar optical and mechanical properties. Gold is a representative of the class of soft (small Young modulus) materials that are also employed in MEMS and NEMS. Another relevant feature of gold is its high optical reflectivity, which results into a stronger Casimir force compared to a semiconductor \cite{GussoJPD}. The results shown in  Fig. \ref{contour} are for resonators with a constant aspect ratio $l/w=5$, which is large enough to be representative of a wide number of practical resonators \cite{Nguyen}, but still sufficiently small for the point source approximation to apply. Three values of $Q$ were chosen, encompassing values that are currently obtained for micro- and nanoresonators ($Q = 10^4$ and $10^5$) and those expected from technological improvements in future devices ($Q = 10^6$).

What is revealed by Fig. \ref{contour} and other similar analysis we performed is that the new energy loss mechanism can be expected to be more relevant when the gaps involved are smaller than approximately 10 nm. Due to the strong dependence of $Q$ on the gap distance, this result holds also  for structures with a thickness considerably larger than the one we considered in Fig. \ref{contour} ($h = 0.1 \, \mu$m), since a small decrease in the gap suffices to compensate for large changes in $h$. Because the predicted quality factor is smaller for small gaps and  thin resonators, this mechanism should be more relevant for nanoresonators actuated electrostatically, since in this case nanogaps would arise naturally. In fact, as mentioned in Sec. \ref{Introduction},  electrode-to-resonator gaps as small as 20 nm were successfully  fabricated for electrostatic actuation and readout of RF signals using a blade nanoresonator. It is worth to mention that this nanoresonator was very long ($ l = 30 \, \mu$m) and had a base thickness of about 1 $\mu$m, demonstrating that tiny gaps can be produced for comparatively large (order o micrometer) structures, and indicating that sub-10 nm gaps can be a common feature in near future micro- and nanoresonators.

Due to the fact that the resonator mechanical energy is transferred over a vacuum gap and dissipated into a nearby structure,  this energy loss can be considered as an example of noncontact friction \cite{Volokitin}. Noncontact friction  has been considered so far mainly in the context of noncontact atomic force microscopy (nc-AFM) \cite{Bhushan}, and several mechanisms involving the tip-sample interaction were  considered to be the source of the energy dissipation. Acoustic energy loss was  considered in this context \cite{Volokitin}, but the results derived so far are valid for small tip oscillations (harmonic approximation) and can not be compared with, for instance, the precise experimental data presented in Ref. \onlinecite{Gostmann}, measured using large amplitude nc-AFM. Other possible sources of noncontact friction are anelastic processes on the tip and sample, van der Waals friction resulting from fluctuating electromagnetic field \cite{Explanation}, and electrostatic friction involving electromagnetic emissions or the Joule effect. However, the bulk of experimental data can not be consistently explained by any one of the  known noncontact friction mechanisms \cite{Volokitin,Bhushan,Hoffmann,Pfeiffer,Garcia}, instead specific models (usually purely phenomenological models) of energy dissipation were used to explain the data for each experiment. However, some of these noncontact friction mechanisms could also contribute significantly to the total dissipation of micro- and nanoresonators.  

An approximate but  straightforward comparison between some of  the different noncontact energy loss mechanisms can be done by comparing the friction coefficient per unit of area resulting for each mechanism. This coefficient can be obtained by modelling the resonator as an one degree of freedom system subject to a viscous damping.  For the first mode of both cantilever and bridge resonators the coefficient of friction due to the acoustic losses is given by $\gamma = m_{eff} \, \omega_1/(Q S)$, where $m_{eff} = c \, \rho \, h w l$ denotes the effective mass, with $c=0.396$ for bridge and $c=0.250$ for cantilever,  $Q$ corresponds to the quality factor and $S=w l$ to the area. The resulting $\gamma$ is proportional to the area, and for a microresonator made from gold with $l=5w= 5 \mu$m suspended 10 nm above a gold substrate it results to be $\gamma = 0.32$ Kg s$^{-1}$ m$^{-2}$.  For comparison, the corresponding  friction coefficient due to the van der Waals friction assuming clean gold surfaces is approximately $\gamma^{vdW}=10^{-5}$ Kg s$^{-1}$ m$^{-2}$ at a temperature $T=300$ K \cite{Volokitin}. For semiconductors $\gamma^{vdW}$ increases, and for silicon carbide it is predicted to be one order of magnitude larger than for good conductors. Almost exactly the opposite of that is observed for $\gamma$ which is approximately one order of magnitude smaller for a semiconductor like silicon as compared to gold.  It is worth to mention that the van der Waals friction is expected to increase by orders of magnitude under certain circumstances \cite{Volokitin}, for instance, with surface contamination, therefore giving rise to a significant energy loss channel, possibly comparing to or surpassing the energy loss mechanism analyzed in this work. It is also interesting to note that the Joule dissipation \cite{Joule}, due to the Joule effect, investigated in the context of nc-AFM, has already been  incorporated into the modelling of practical micro- and nanoresonators actuated electrostatically. In this last case the time varying electric field, due to resonator vibration, results into a time varying charge at the electrodes and, consequently, an electric current. The vibrational energy is dissipated  by the Joule effect as this current flows through the structure facing the electric resistance $R$ forming the equivalent RLC circuit \cite{delosSantos}.

We conclude by noting that the energy loss mechanism we investigated can be relevant for a wide class of future NEMS and MEMS where moving  parts are separated by distances at the nanoscale. The ubiquitous Casimir force can produce the coupling between the moving parts and nearby structures through which mechanical energy can be lost in ways that were not addressed in this work. Therefore, further investigations on the implications of this energy loss mechanism on different systems should be performed.

\begin{acknowledgments}
The author acknowledges the financial support by the  Conselho Nacional de Desenvolvimento Cient\'ifico e Tecnol\'ogico, CNPq-Brazil.
\end{acknowledgments}


\begin{thebibliography}{}
\bibitem{Explanation} The van der Waals friction should not be confused with the acoustic energy loss mediated by Casimir/van der Waals force we investigate in this work. The van der Waals friction has a physical origin similar to that of Casimir/van der Waals force, the fluctuations of the electromagnetic field. However, van der Waals friction is mostly determined by the exchange of Doppler-shifted real photons due to the relative motion of the bodies \cite{Volokitin}.
\end{thebibliography}
\end{document}